\documentstyle[prl,twocolumn,aps,tighten,floats,epsfig]{revtex}
\begin{document}
\draft
\twocolumn[\hsize\textwidth\columnwidth\hsize\csname @twocolumnfalse\endcsname
\title{Re-integerization of fractional charges in the correlated quarter-filled band}

\author{R.T. Clay$^{1,2}$, S.~Mazumdar$^1$, and D.K.~Campbell$^3$}

\address{$^1$ Department of Physics, University of Arizona, Tucson, AZ 85721}
\address{$^2$ Cooperative Excitation Project ERATO, Japan Science and 
Technology Corporation (JST), University of Arizona, Tucson, AZ 85721}
\address{$^3$ College of Engineering, Departments of Electrical and Computer 
Engineering
and Physics, 
Boston University, Boston, MA 02215}
\date{\today}
\maketitle
\begin{abstract}
Previous work has demonstrated the existence of soliton defect
states with charges $\pm$ e/2 in the limits of zero and of infinite on-site Coulomb 
interactions in the one-dimensional (1D) quarter-filled band.
For large but finite on-site Coulomb interaction, the low temperature 2k$_F$
bond distortion that occurs within the 4k$_F$ bond-distorted phase
is accompanied by charge-ordering on the sites. We show that a ``re-integerization''
of the defect charge occurs in this bond-charge density wave (BCDW) state  
due to a ``binding'' of the fractional charges.
We indicate briefly possible implications of this result for mechanisms of organic 
superconductivity.
\medskip

\noindent PACS indices: 71.30.+h, 71.38.+i, 71.45.Lr, 74.70.Kn  
\end{abstract}]

\narrowtext
One of the most dramatic and exciting predictions of the initial model
studies of quasi-1D organic $\pi$-conjugated polymers and charge transfer
solids (CTS) was the existence of exotic excitations, including solitons with 
unusual spin-charge relations \cite{review,Rice1} and fractional charge 
\cite{Su,Ono1,Ono2,Rice,Howard,ZKG}. 
For trans-polyacetylene, with a 1/2-filled 1D band, it was shown within the
Su-Schrieffer-Heeger (SSH) one-electron model that the neutral solitons have
spin 1/2, while charged solitons are spinless \cite{review,Rice1}. 
This result remains
unaltered upon inclusion of the Coulomb electron-electron (e-e) interactions
\cite{Hirsch,DeGrand}.
In the case of the 1D 1/3- and 1/4-filled bands, any non-zero electron-phonon (e-ph)coupling leads to 2k$_F$ Peierls transitions 
corresponding to trimerization and tetramerization of the respective lattices, and
defect soliton charges are $\pm$ 2e/3 \cite{Su,Ono1} and $\pm$ e/2 \cite{Ono2}, 
respectively.
Effects of e-e interactions on the soliton charge in non-1/2-filled bands
have been investigated only for the 1/4-filled band, 
primarily for the limiting case of 
large on-site Hubbard repulsion ($U \to \infty$)  
\cite{Rice,Howard,ZKG}.
For $U \to \infty$, the
1D 1/4-filled Hubbard band of electrons reduces to a 1/2-filled band of 
{\it noninteracting} spinless 
fermions \cite{Rice,Howard,ZKG}. Hence in the presence of an e-ph coupling, a 
Peierls transition occurs at twice the fermi wave vector corresponding
to the spinless fermions, which is four times the original k$_F$
(k$_F$ = $\pi$/4a, where a is
the undistorted lattice constant).  Depending on the nature of the phonons,
this leads to bond or site {\it dimerization} of the lattice.
A single doped hole/electron generates {\it two} charged soliton defects in the dimerized lattice,  which therefore have charge
$\pm$ e/2 again. Since soliton charges are e/2 at both $U = 0$ and $U \to \infty$,
a continuity between these two limits
has been claimed \cite{ZKG}.

In the case where the 4k$_F$ metal-insulator transition is due to bond distortion,
a 2k$_F$ spin-Peierls (SP) dimerization of the dimerization
({i.e.}, tetramerization) occurs at low temperatures 
for {\it finite} $U$ \cite{SP}. 
Reference \onlinecite{ZKG}, which claimed the persistence of fractional
charge in the 2k$_F$ phase, did not actually probe the nature of the 2k$_F$.
We have recently shown that the 2k$_F$ SP bond tetramerization is also accompanied
by an on-site 2k$_F$ CDW \cite{umt}, and the correct description of this state
is a bond-charge-density wave (BCDW). Clearly an important theoretical question is
whether fractionally charged solitons continue to exist in the BCDW ground state.

This issue is of considerably more than mere
theoretical interest in the case of the organic CTS.
The original applications of the ``large-$U$'' 1/4-filled
band theory were to 1:2 {\it anionic} TCNQ solids \cite{Epstein}, where, 
however, no direct evidence for fractional charge was
found, in spite of experimental attempts. More recently, interest has shifted to 
the nominally 1/4-filled 2:1
{\it cationic} CTS, many of which exhibit superconductivity and other novel phases
at low temperatures \cite{review1}. The most strongly 1D systems here,
(TMTTF)$_2$X (X = PF$_6$, AsF$_6$ etc.) exhibit a high temperature metal-insulator
(presumably 4k$_F$) transition, and a SP transition at T$_{SP} <$ 20 K. Again,
no evidence for fractionally charged defect states has emerged. The question
that arises then is whether the half-integer charges are unobservable, or the
right experiments have not been done yet. This question acquires added importance
when the entire family of 2:1 cationic CTS is considered. 
In our recent theoretical studies, for example, we have shown that the BCDW
persists in the interacting 1/4-filled band for all $t_{\perp} \leq t$, where
$t$ and $t_{\perp}$ are the intra- and interchain one-electron hopping integrals, 
respectively, and furthermore, in the small $t_{\perp}$ region the BCDW 
also coexists with a
spin-density wave (SDW),
giving a BCSDW \cite{PRL,PRB}. 
Experimentally, charge-ordering \cite{Chow,Monceau} has been found
in the 1D (TMTTF)$_2$X which are SP systems, while the BCSDW has been 
seen in (TMTSF)$_2$X 
\cite{Pouget,Kagoshima} and $\alpha$-(BEDT-TTF)$_2$MHg(SCN)$_4$ 
\cite{Toyota}. Evidence for coexisting SP-like states and charge-ordering is also 
beginning to
emerge in other BEDT-TTF systems \cite{PRB,Musfeldt}.
Based on the persistent
BCDW that is obtained theoretically and also observed experimentally,
we have suggested that superconductivity in these strongly correlated
systems is a consequence of pairing of commensurability
defects of the BCDW phase \cite{PRB}. If these commensurability defects are
fractionally charged then, an obvious question that emerges is whether the
the superconducting  ``pairs'' are actually ``quadruplets''! 

In the present Letter, we show that charge-fractionalization {\it does not 
occur} in the low
temperature 2k$_F$ SP phase of the interacting 1D 1/4-filled band:
the charge-ordering
associated with the SP tetramerization causes a ``re-integerization'' 
of the fractionally charged defects
that occur within the 4k$_F$ dimerized phase. 
This re-integerization provides support to the idea that superconductivity
in the quasi-2D TMTSF and BEDT-TTF may be due to pairing of integer charged
commensurability defects in a background BCDW, although it does not necessarily
exclude the ``melting'' of the BCDW at the superconducting transition.

To establish this result, we will consider the 1D Peierls-extended Hubbard 
Hamiltonian
for the case of bandfilling slightly 
away from
1/4, and will present both many-body numerical results as well as heuristic
physical arguments. No SP transition can occur within the 4k$_F$ CDW \cite{PRB},
and therefore the the experimentally observed SP transition in (TMTTF)$_2$X proves 
that the 4k$_F$ metal-insulator transition is to a bond-order wave (BOW) 
that occurs for
intersite e-e interaction $V < V_c \sim 2|t|$ \cite{PRB}.
Upon the exclusion of the 4k$_F$ charge-ordering, there
are four possible equivalent
charge-orderings that can occur for nonzero $U$ and $V < V_c$, viz., ...1100...,
...0110..., ...0011..., and ...1001... (here and below a `1' implies charge
0.5 + $\epsilon$, and a `0' implies 0.5 -- $\epsilon$).
The 4k$_F$ bond dimerization breaks a twofold
symmetry and the ground state is a superposition of the type
...1100... + ...0011..., with weak ($W$) `11' and `00' bonds, and strong ($S$)
`10' and `01'
bonds. The 2k$_F$ SP transition breaks yet another twofold symmetry, and the
ground state is now {\it either} ...1100... {\it or} ...0011..., with the 
`11' ($W'$)
and `00' ($W$) bonds now inequivalent, and the bond distortion pattern $W'SWS$
\cite{umt,PRL,PRB}.
Our goal is to calculate
defect charges within this $W'SWS$ state.

Since our many-body numerical results are for finite chains, 
it is useful
to establish our methodology by first solving the known case of
$U \to \infty$, $V$ = 0. The spinless fermion
Hamiltonian here is written as,
\begin{eqnarray}
H_{SF} = -t\sum_j (a^\dagger_ja_{j+1}+h.c)
\label{spinless}
\end{eqnarray}
For simplicity,
we have not included the SSH e-ph interaction that gives the bond
dimerization in the above \cite{Rice,ZKG}. As we show below,
calculations of the bond orders, defined as
$\langle B_{j,j+1} \rangle$ = $\langle a^\dagger_ia_{i+1}+h.c\rangle$, 
and of site charge densities,
of an open long finite chain with 
uniform hopping $t$ show {\it spontaneous} modulations of the bond orders and
charge densities,
with the {\it same} modulation patterns as expected within a fully
self-consistent calculation of a periodic electron-phonon coupled ring.
The actual lattice dimerization occurs for 0$^+$ electron-phonon
coupling in the infinite chain. 

\begin{figure}[htb]
\centerline{\epsfig{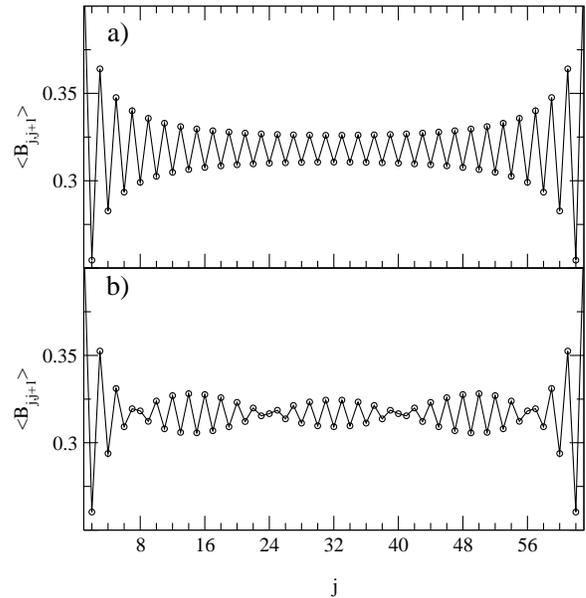}}
\medskip
\caption{The bond-orders of a 64-site uniform open chain of 
noninteracting spinless fermions with (a) 32 fermions, showing the 
spontaneous dimerization
that occurs for the effective half-filled band; 
(b) two doped holes (30 fermions),
showing the resulting {\it four} defects.}
\label{spinless-64}
\end{figure}

Figure~\ref{spinless-64}(a) shows the bond orders of a 64-site open chain of
noninteracting spinless fermions with uniform hopping integrals; 
the spontaneous period-two dimerization of the
chain is evident in spite of end effects.
In Figure~\ref{spinless-64}(b), we show the result of adding 
two holes. 
The two holes produce four defects (regions
where consecutive bond orders are equal), each of which
then have charge + e/2. Charge density calculations (not shown) 
give uniform $\langle n_j \rangle$ = 0.5 for the undoped chains, and four defects
in the doped chain, with maximum deviations from 0.5 ocurring exactly in 
the regions where consecutive bond orders are equal.

\begin{figure}[htb]
\centerline{\epsfig{width=3.0in,file=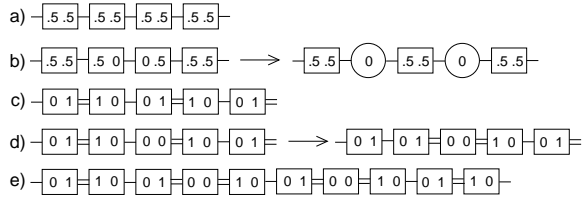}}
\medskip
\caption{(a) The undoped 4k$_F$ dimerized 1/4-filled band. 
Boxes represent the ``dimers'', {\it i.e},
sites coupled by strong bonds.
(b) creation of two e/2 solitons upon the removal of one electron
from the configuration of (a).
(c) The BCDW state that describes the
interacting 1/4-filled band, below the 2k$_F$ SP transition (see text).
(d) A doped hole removes a single `1', and soliton charge is now e.
(e) Configuration
with two doped holes in the 2k$_F$ BCDW.}
\label{cartoon}
\end{figure}

Although the above figures reproduce known results, it is useful to give a 
configuration space interpretation of these results.
In Figure~\ref{cartoon}(a) we show the spontaneously dimerized ground state, 
with equal  
$\langle n_j \rangle = 0.5$ on
each site. Here the bonds between the sites within a box are stronger than the
bonds between sites that occur in neighboring boxes. 
Removal of a single electron initially removes two
charges 0.5 from each of two adjacent sites connected by a weak bond 
(see first diagram, Figure~\ref{cartoon} (b)).
The configuration with adjacent defect sites has higher energy 
than the one in which they are separated \cite{review,Rice1}, which leads to a
spontaneous separation of the defects (see second diagram, 
Figure~\ref{cartoon} (b)),
each carrying charge e/2.
The defects are shown localized to a single site in Fig.~\ref{cartoon}(b) 
for simplicity; as seen in the numerical data of
Fig.~\ref{spinless-64}(b), 
each defect 
actually spreads
out over several sites.

To explore the
defect charges for finite $U$ below the 2k$_F$ transition,
we study the Hamiltonian \cite{PRB,Ogata}:
\begin{eqnarray}
H &=& - t_1 \sum_{j,\sigma} B_{2j-1,2j,\sigma} 
 - \sum_{j,\sigma}[t_2 - \alpha \Delta_j] B_{2j,2j+1,\sigma} \\
& & + {K\over2}\sum_{j} \Delta_j^2 
 + U\sum_{j} n_{j,\uparrow}n_{j,\downarrow} + V\sum_{j}n_{j} n_{j+1}. \nonumber
\label{hamiltonian}
\end{eqnarray}
In the above  
$B_{i,j,\sigma}=c^\dagger_{i,\sigma}c_{j,\sigma}+h.c.$ is
the standard hopping operator for an electron with spin $\sigma$,
$n_{j,\sigma}=c^\dagger_{j,\sigma} c_{j,\sigma}$ and $n_j= \sum_{\sigma}
n_{j,\sigma}$
the standard electron-number operators, 
$\alpha$ and $K$ the electron-phonon coupling and the spring constants, 
respectively,
and $\Delta_j$
is the displacement of each dimer
from equilibrium. The intra-dimer hopping $t_1$ is larger than
the inter-dimer hopping $t_2$ 
because of the 4k$_F$ bond dimerization which is assumed to have 
occurred already.

Before showing our numerical data, we present physical arguments based on
a configuration space picture that already suggest
re-integerization of the fractional charge within the above Hamiltonian. 
As shown in our previous work (see also below), the ground state of the
commensurate interacting 1/4-filled band is the BCDW state shown in 
Figure~\ref{cartoon}(c). Here each `10' and `01' within a box are connected
by the strong (S) bond, while the ``double'' and ``single'' bonds correspond
to the $W'$ and $W$ bonds, respectively. 
The ...1100... charge-ordering dictates that ``doping'' now will involve removal
of a whole electron from a single site, as shown in the first diagram of
Figure~\ref{cartoon}(d). Simply removing the `1' creates an 
energetically
unfavorable asymmetric situation 
with three `0's in a row, and energy is
gained by the reorganization shown in the second step of 
Figure~\ref{cartoon}(d), 
which reduces the number of weak `0-0' bonds.
The difference from the $U \to \infty$ case is then as follows. The
defect induced charge deficiency now occupies a single box 
and no further separation between the charges can occur. Clearly,
the soliton charge is then +e and not +e/2. Note that the charge reorganization
creates phases of inter-dimer bond dimerizations that are different on the
left and the right of the soliton defect (this may be seen most easily by
numbering the boxes sequentially).
Soliton defects with integer charge therefore occur in
pairs, where a second hole ``heals'' the ``phase-problem'' induced by the first
hole, as shown in Figure \ref{cartoon}(e). 
Note that the above physical picture, though qualitative, predicts
two important numerically testable details,
viz., (i) the intra-dimer bonds become weak near the defect center, 
with a minimum in the
intra-dimer bond order occurring over a single dimer, (ii) the inter-dimer
bonds around a defect {\it increase} in
strength,  
and reach
their maximum values on both sides of the weakest intra-dimer bond. 
 
\begin{figure}[htb]
\centerline{\epsfig{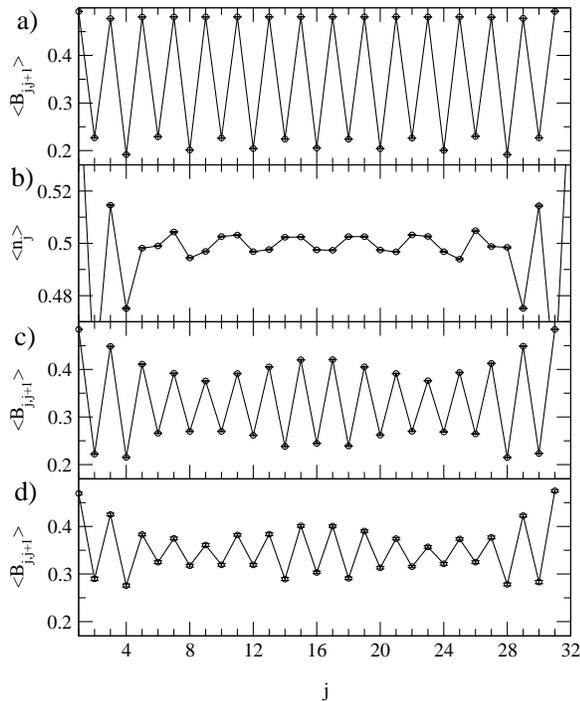}}
\caption{SSE bond orders of
32 site open chains with $U=6$,
$V=1$. The lines are guides to the eye.
(a) Undoped dimerized 1/4-filled system, $t_1=1.4$ and $t_2=0.6$.
Spontaneous dimerization of the dimer lattice results in the alternation of
weaker bond orders.
(b) Charge densities of the undoped chain, showing ...1100... ordering.
(c) Bond orders with two added holes, with the same parameters as in (a).
(d) Same as (c), but with weaker intrinsic dimerization, $t_1=1.2$
and $t_2=0.8$.}
\label{data-32}
\end{figure}

In our numerical investigation we study the 
bond orders $\sum_{\sigma}\langle B_{j,j+1,\sigma}\rangle$ and
charge density $\sum_{\sigma} \langle n_{j,\sigma} \rangle$, 
at and near 1/4-filling.
We use the Stochastic
Series Expansion (SSE) quantum Monte Carlo (QMC) method \cite{SSE}, which 
is exact to within statistical error for this 1D
system: there is no approximation in the discretization of
imaginary time, as in many other QMC methods. As SSE is a
finite-temperature QMC method, a sufficiently low temperature
must be used to obtain ground-state results. In the results
shown below, inverse temperatures $\beta=4L$ ($L$=chain length) were used, 
which showed essentially no difference from
$\beta=2L$ computations. Since bond-distortions in open chains {\it decrease} with 
length, unlike in periodic rings, where they increase (convergence to the same 
finite amplitude for the BOW occurs in the long chain limit)  
our initial calculations are
for relatively short chains with 32 sites.
Figure~\ref{data-32}(a) shows the bond orders for a 32-site
open chain with 16 electrons
and $U=6$, $V=1$, and hopping integrals $t_1=1.4$
and $t_2=0.6$ (in units of $t = (t_1+t_2)/2$). 
It is seen that the strong intra-dimer bonds 
have nearly uniform
bond-orders, but there occurs a {\it spontaneous alternation of the inter-dimer bond
orders}, 
giving an overall bond distortion pattern  
$W'SWS$.  
Figure~\ref{data-32}(b) shows the charge densities in this
undoped 1/4-filled system, and there is clear spontaneous site-CDW with
periodicity ...1100...  
in the center of the chain, with relatively weak chain end effects. 
These results show that once the 4k$_F$ bond
dimerization ($t_1 \neq t_2$) is explicitly put in 
``by hand'',
further dimerization of the dimerization is unconditional. 

In Figure~\ref{data-32}(c) we show the bond orders for the dimer chain ``doped''
with two holes.
There occur now only two defects, 
centered on the dimer
bonds between sites 9 and 10, and sites 23 and 24, respectively. These
defects therefore have integer charges. Note that
the intradimer bond orders are smallest for the 9--10 and 23--24 bonds,
while the interdimer ``weak'' bonds on both sides of the defect centers 
reach their
strongest values, in complete agreement with 
Figures~\ref{cartoon}(d) and (e).

The very large difference between $t_1$ and $t_2$ in 
Figures~\ref{data-32}(a) - (c)
was chosen merely to make the results visually most accessible, 
and is not a required
condition for obtaining defects with integer charge.
To demonstrate this, we have performed calculations with smaller $t_1 - t_2$. 
In Figure~\ref{data-32}(d) we show our results for the same 32-site system with
two doped holes, but now for weaker 4k$_F$ dimerization
$t_1=1.2$ and $t_2=0.8$. No change in the defect
structure is seen with the weaker dimerization. 

\begin{figure}[htb]
\centerline{\epsfig{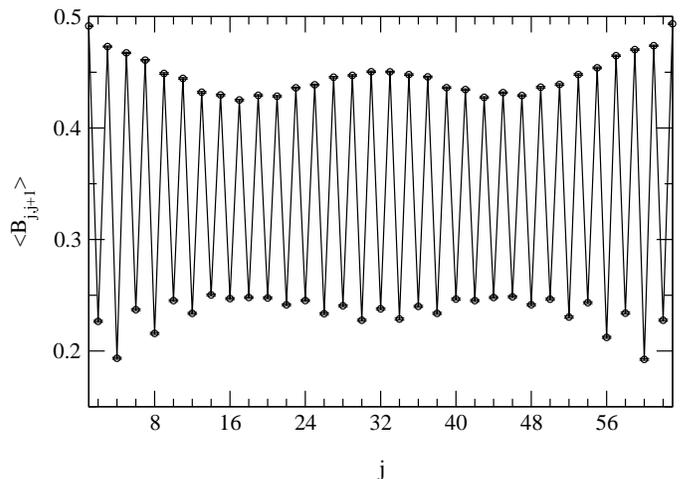}}
\caption{SSE bond orders of 
a 64 site chain with two added holes, for $U=6$,
$V=1$, $t_1=1.4$, $t_2=0.6$.}
\label{data-64}
\end{figure}

Finally, 
it is important to demonstrate that the occurrence of only two defects in 
Figures~\ref{data-32}(c) and (d) is not a consequence of short chain lengths 
and
overlapping defects. We have therefore performed our calculations also for 64 sites,
a chain length where $U \to \infty$ shows clear evidence for four defects
(see Figure~\ref{spinless-64}). Based on the similarity between large and
moderate $t_1 - t_2$ in Figures.~\ref{data-32}(c) and (d), 
we performed our calculations
here for only large $t_1 - t_2$, viz., $t_1$ = 1.4, $t_2$ = 0.6. 
Our SSE results for a 64-site open chain with 30 spinful electrons
are shown in Figure~\ref{data-64}, where once again only two defects 
(centered on the
dimer bonds between sites 17 and 18, and sites 45 and 46, respectively \cite{note}) 
are seen. An important feature that is clearer in this longer chain
is that the weaker bond orders correctly recover their strong-weak character away 
from the defect centers.

To conclude, the apparent continuity of fractional charges between the $U=0$
and $U \to \infty$ limits in the 1/4-filled band \cite{ZKG} arises only when 
the existence of the 2k$_F$ BCDW ground state at finite $U$ is ignored.
The charge e/2 solitons that occur  
at $U=0$ are domain walls between four distinct density wave phases, while
at $U \to \infty$ they are domain walls between only two such phases. The
e/2 charge at $U \to \infty$ is thus a consequence of site-occupancies of
0.5 electrons alone, and re-integerization
of the fractional charge occurs in the 2k$_F$ BCDW ground state. The integer
charge solitons do not bind in 1D, but it is conceivable that integer charge 
commensurability defects in the 2D BCDW bind into large bipolarons, thereby
giving the superconductivity observed in the 2D CTS. Recent demonstrations
of large bipolarons \cite{Aubry} with strongly interacting 2D electrons,
and the finite mobility of such intersite bipolarons \cite{Aubry,Trugman} are
highly promising results in these context. 

RTC acknowledges helpful discussions with A.W. Sandvik regarding
the SSE method. DKC acknowledges support from NSF DMR-97-12765.
Numerical calculations were done in part at the 
NCSA.


\begin{references}
\bibitem{note} From symmetry, the defect on the right should be centered on the
bond between sites 47 and 48. As seen in Figure \ref{data-64}, the bond orders
for the bonds 45-46 and 47-48 are nearly the
same. The inter-dimer bonds, are, however, equal for the two bonds on both sides
of the 45-46 bond, and this is why we believe that the defect center is located on
this bond. This small discrepancy is a consequence of statistical errors within the
QMC method.
\end{references}
\end{document}